\documentclass{article}
\usepackage{spconf,amsmath,graphicx}
\usepackage{multirow}
\usepackage{xcolor}
\usepackage{booktabs}
\usepackage{algorithm}
\usepackage[hidelinks]{hyperref}
\usepackage{placeins}
\usepackage{colortbl}
\usepackage[belowskip=-12pt,aboveskip=0pt]{caption}

\definecolor{tueScharlaken}{HTML}{C81919} 
\definecolor{tueblue}{rgb}{51,51,178} 
\newcommand{\gr}{\color{gray}}
\newcommand{\red}{\color{tueScharlaken}}

\def\spacebeforesecs{-2mm}

\title{Spectral Masking with Explicit Time-Context Windowing for Neural Network-Based Monaural Speech Enhancement}
\name{%
\begin{tabular}{@{}c@{}}
Luan Vinícius Fiorio$^{\star\dagger}$, 
Boris Karanov$^{\star\dagger}$, 
Bruno Defraene$^{\dagger}$, 
Johan David$^{\dagger}$, \\
Wim van Houtum$^{\star\dagger}$,
Frans Widdershoven$^{\dagger}$, 
Ronald M. Aarts$^{\star}$
\end{tabular}}

\address{$^{\star}$ Eindhoven University of Technology, Eindhoven 5600 MB, The Netherlands \\
$^{\dagger}$ NXP Semiconductors, High Tech Campus 46, Eindhoven 5656 AE, The Netherlands}

\begin{document}
\maketitle
\begin{abstract}

We propose and analyze the use of an explicit time-context window for neural network-based spectral masking speech enhancement to leverage signal context dependencies between neighboring frames. In particular, we concentrate on soft masking and loss computed on the time-frequency representation of the reconstructed speech. We show that the application of a time-context windowing function at both input and output of the neural network model improves the soft mask estimation process by combining multiple estimates taken from different contexts. The proposed approach is only applied as post-optimization in inference mode, not requiring additional layers or special training for the neural network model. Our results show that the method consistently increases both intelligibility and signal quality of the denoised speech, as demonstrated for two classes of convolutional-based speech enhancement models. Importantly, the proposed method requires only a negligible ($\leq 1$\%) increase in the number of model parameters, making it suitable for hardware-constrained applications.

\end{abstract}


\vspace{\spacebeforesecs}
\section{INTRODUCTION}
\label{sec:intro}


Monaural speech enhancement (SE) often concentrates on the removal of unwanted background noise from a single-channel noisy speech signal \cite{Loizou2013Speech}. Low-complexity SE solutions might employ an acoustic environment classification scheme \cite{Yellamsetty2021Comparison, Fiorio2023Semisupervised} and use a noise reduction technique tailored to the specific noise type. On the other hand, considering various noise types, neural networks (NN) and deep learning (DL) has been successfully applied for carrying out the SE task \cite{Wang2014OnTraining,Tan2018Convolutional,Ullah2022EndToEnd}. Typically, DL-based SE is implemented via supervised learning, which relies on targets such as ideal binary, ideal ratio masks or clean speech spectra \cite{Wang2014OnTraining,Wang2005}. The application of DL can lead to improved performance in terms of speech intelligibility and quality metrics when compared to classical signal processing techniques for noise reduction, such as Wiener filtering \cite{Loizou2013Speech}.

Supervised DL-based SE can operate either in time-domain or time-frequency (T-F) domain. While time-domain processing has attracted attention in recent years \cite{Pandey2019TCNN, Kong2022Speech}, traditionally T-F features have been used for SE due to well established hardware implementations of the fast Fourier transform and its inverse operation \cite{Benesty2011Speech}. In particular, enhancement of the magnitude of T-F features is the lowest complexity processing, which however results in a suboptimal solution as it does not process the noisy phase. Typically, state-of-the-art magnitude processing SE deep learning solutions are based on convolutional models \cite{Tan2018Convolutional, Ullah2022EndToEnd, Pandey2019TCNN, Kong2022Speech, Shahriyar2019, Tan2020Learning}, which are often characterized by fewer parameters compared to their fully connected and recurrent counterparts.

To increase the speech enhancement quality of NN-based systems, a common approach is to increase the number of layers, increase the layer dimensions, or resort to layers with higher modeling capacity, e.g., transformers. This comes with a higher computational footprint in terms of the number of  model parameters, and the number of multiplications required to execute the model. Therefore, in computationally-constrained applications, alternative approaches which do not increase the complexity of the model could be desirable.

In this paper, we propose a \emph{post-processing} method aimed at increasing speech enhancement quality with a limited impact on the NN topology and hence on the computational footprint. More specifically, targeting lower complexity implementations, we develop the method for T-F magnitude-based speech enhancement. Nevertheless, as discussed in the manuscript, our approach is general and can be applied in different scenarios. In particular, we investigate the use of an explicit time-context window for NN-based SE for post-processing. The approach recombines the estimated soft mask values in multiple time-dependent contexts via a sliding window technique, using the average operation. Differently, while DeepFilterNet \cite{schroter2022deepfilternet} also combines multiple estimations in different contexts, the proposed algorithm doesn't require additional layers to estimate the combination weights since it's defined as a post-processing inference approach.

Our results show that, for the wide range of considered noise types and signal-to-noise ratios (SNR), the sliding window post-processing achieves consistent improvements both in the short-time objective intelligibility (STOI) and perceptual evaluation of speech quality (PESQ) metrics for convolutional-recurrent network (CRN) and convolutional denoising autoencoder (CDAE) models.

\vspace{\spacebeforesecs}
\section{Preliminaries}

\label{sec:preliminaries}

Let $S(t,f)$, $N(t,f)$, and $Y(t,f)$ be the short-term Fourier transform (STFT) representations of the clean speech signal $s(t)$, noise signal $n(t)$, and noisy speech signal $y(t)$, where $t$ is the discrete time index, $f$ is the discrete frequency index, and $Y(t,f) = S(t,f) + N(t,f)$. The ideal ratio mask (IRM) is then defined as \cite{Wang2014OnTraining}
\begin{equation}
    M(t,f) = \left( \frac{|S(t,f)|^2}{|S(t,f)|^2 + |N(t,f)|^2} \right)^{\beta},
\label{eq:IRM}
\end{equation}
with a compression parameter $\beta \in (0,1]$. The denoised speech STFT can be obtained by
\begin{equation}
    \hat{S}(t,f) = M(t,f) \odot Y(t,f),
\label{eq:Shat}
\end{equation}
where $\odot$ represents the Hadamard product. Notice that \eqref{eq:Shat} multiplies a real-valued matrix -- $M(t,f)$ -- with a complex-valued matrix -- $Y(t,f)$. Thus, only the magnitude of $Y(t,f)$ is denoised, while the phase is kept the same. Speech denoising systems aim to provide an IRM estimate $\hat{M}(t,f)$, and reconstruct the clean speech STFT by $\hat{S}(t,f) = \hat{M}(t,f) \odot Y(t,f)$.

\vspace{\spacebeforesecs} 
\begin{figure}[!t]
    \centering
    \includegraphics[width=0.47\textwidth]{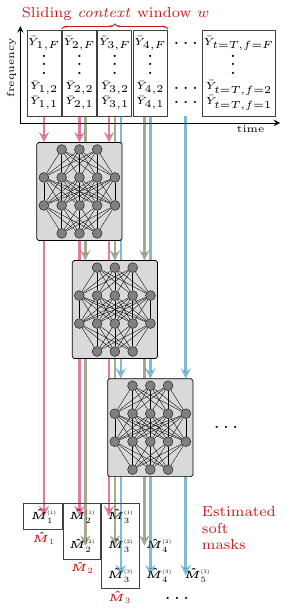}
    \caption{Sliding context window schematic example for $w=3$.}
    \label{fig:sliding_window}
\end{figure}

\section{Sliding window post-processing}

\label{sec:sliding}

From a communication system perspective, it has been shown that a sliding window technique combined with simple recurrent neural networks leads to close-to-optimal sequence detection \cite{Farsad2018Neural, Karanov2019EndToEnd, Karanov2020}. In \cite{Wang2014OnTraining}, it has been hinted that additional temporal dependencies can be exploited via windowing operations. From that perspective, in this section we devise a post-processing framework for enhanced estimation of STFT masks in presence of temporal dependencies.

Figure~\ref{fig:sliding_window}, adapted from \cite{Karanov2019EndToEnd}, depicts the sliding window process which we apply to the problem of speech enhancement. The next subsection describes the method in more detail.

\subsection{Method description}

The sliding window scheme for the neural network processor operates on the time axis of the time-frequency bin features of the noisy speech. It estimates the IRM for $w$ bins and then slides by one bin. The resulting multi-context estimates of a mask at time $t$ are combined to obtain a refined final mask.

\begin{figure*}[!ht]
    \centering
    \includegraphics[width=1.0\textwidth]{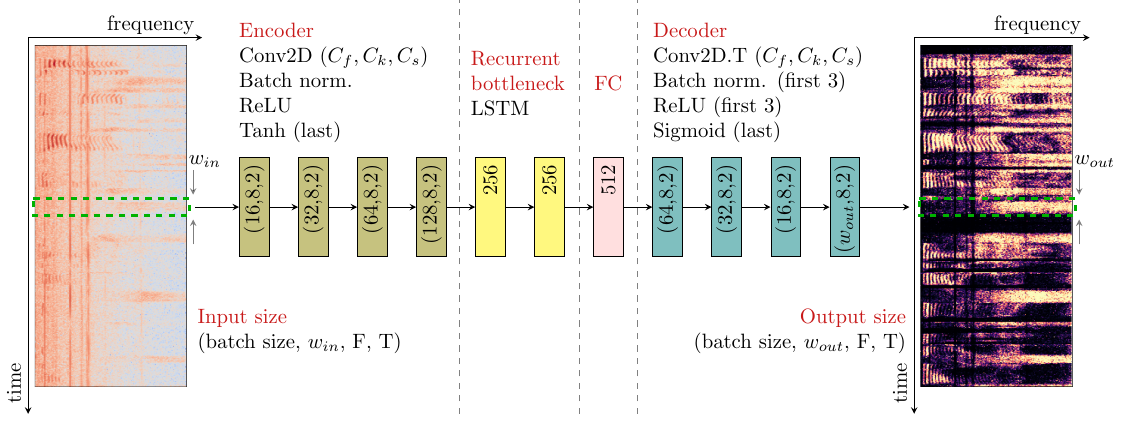}
    \caption{Schematic of the considered convolutional recurrent neural network.}
    \label{fig:CRN}
\end{figure*}

\sloppy Given an STFT sequence duration $T$, the magnitude of the normalized -- by it's mean and standard deviation -- noisy speech STFT is denoted by the matrix $\bar{Y} = [\bar{\boldsymbol{Y}}_{1} \ \bar{\boldsymbol{Y}}_{2} \ ... \ \bar{\boldsymbol{Y}}_{T}]$, where $\bar{\boldsymbol{Y}}_{t}$ are vectors of $F$ frequency features of the noisy signal at time bin $t$ with STFT frames spaced apart by a hop duration of $t_{hop}$. The NN model has as inputs a context window of $w$ time bins $[\bar{\boldsymbol{Y}}_{t-w+1}\ ...\ \bar{\boldsymbol{Y}}_{t}]$, estimating the corresponding $w$ time bins of the ratio mask at the output $[\boldsymbol{\hat{M}}^{(t-w+1)}_{t-w+1}\ ...\ \boldsymbol{\hat{M}}^{(t-w+1)}_{t}]$. The position of the window is then shifted by one time bin, with new input window $[\bar{\boldsymbol{Y}}_{t-w+2}\ ...\ \bar{\boldsymbol{Y}}_{t+1}]$ and output estimates $[\boldsymbol{\hat{M}}^{(t-w+2)}_{t-w+2}\ ...\ \boldsymbol{\hat{M}}^{(t-w+2)}_{t+1}]$. The refined mask $\hat{M}_t$ is estimated by averaging masks estimated from different time context, as follows. For the first $w-1$ time frames of the IRM will be estimated as \cite{Karanov2019EndToEnd}
\begin{equation}
    \boldsymbol{\hat{M}}_{t} = \frac{1}{t} \sum_{k=1}^{t} \boldsymbol{\hat{M}}^{(k)}_{t}, \quad t=1,\ ...,\ w-1,
\label{eq:sliding_window_first_bins}
\end{equation}
where $k$ is the (shift) iteration step of the IRM estimation. For $w \leq t \leq T-w+1$, the estimation of $\boldsymbol{\hat{M}}$ becomes
\begin{equation}
    \boldsymbol{\hat{M}}_{t} = \frac{1}{w} \sum_{k=t-w+1}^{t} \boldsymbol{\hat{M}}^{(k)}_{t}, \quad t=w,\ ...,\ T-w+1.
\end{equation}
Finally, the last $w-1$ time bins are estimated according to
\begin{equation}
    \boldsymbol{\hat{M}}_{t} = \frac{1}{T-t+1} \sum_{k=t-w+1}^{T-w+1} \boldsymbol{\hat{M}}^{(k)}_{t}, \ \ t=T-w+2,\ ...,\ T.
\end{equation}
Notice that the sliding window method is causal, since the time context at the input of the neural network is composed of $w-1$ past and the current bin.

Differently to the DeepFilterNet \cite{schroter2022deepfilternet} approach, where the network estimates coefficients for the combination of multiple clean speech estimations in different contexts (windows), our method is solely applied post-optimization in inference mode since no weight estimation is necessary for the combination of contexts. We combine the time-frequency bins by taking their mean value over different windows. This has the advantage of not requiring additional layers in the neural network model and thus lowers its complexity, making it suitable for hardware-constrained applications.

\subsection{Latency}

A window of time bins as the input for deep neural network speech enhancement has been previously utilized, for example in \cite{Wang2014OnTraining, Tan2018Convolutional}. For real-time applications it is important to mention that this windowing increases the processing latency. The additional latency $t_\ell$ can be calculated as $t_\ell = (w-1) t_{hop}$, where $t_{hop}$ is the STFT frame hop duration in seconds. Notice that the additional latency is fixed for the whole process.

\vspace{\spacebeforesecs}
\section{Speech enhancement neural network}

\label{sec:proposed}

To verify our method, we apply it to two different neural network models. First, we adapt the convolutional recurrent network (CRN) model proposed in \cite{Tan2018Convolutional}, making it more suitable for constrained hardware implementation. Additionally, we consider another model, derived from the CRN, which is representative of a more general class of convolutional autoencoders, resulting in an architecture termed convolutional denoising autoencoder (CDAE) in \cite{Shahriyar2019}.

\subsection{Models description}
\label{sec:models}

The first model \cite{Tan2018Convolutional} is a CRN composed of a convolutional encoder, recurrent bottleneck, and a convolutional decoder. The considered implementation is shown in Figure~\ref{fig:CRN}, where $w_{in}$ and $w_{out}$ represent the input and output context window size, respectively. The convolutions are defined to only operate in the frequency dimension; the total number of layers are reduced in the encoder and decoder of the original model, also removing the memory-expensive skip connections; the long short-term memory (LSTM) layers are also reduced; the activation functions are changed from exponential linear unit to the simpler rectified linear unit (ReLU). This model contains approximately 1.62M parameters. In the figure, FC denotes a fully connected layer, Conv2D denotes 2D convolutional layers, Conv2D.T is the transposed 2D convolution, and $C_f$, $C_k$, and $C_s$ are, respectively, number of feature maps, kernel size, and stride.

Furthermore, to validate the sliding window post-processing for a different model class, the CDAE architecture \cite{Shahriyar2019} which contains approximately only 173k parameters, is considered for a very low complexity study case. The CDAE is a convolutional autoencoder which, differently from the CRN, does not have a recurrent bottleneck based on LSTM and fully connected layer. For the purposes of a fair comparison, we kept the other parameters in the two considered architectures identical, as shown in Figure~\ref{fig:CRN}.

The models estimate the soft masks for $w_{in}$ consecutive frames, resulting in an output of size $(w_{out}, F, T)$, where batch size is omitted for simplicity. For a fair comparison, we also consider modified versions of the models where their output sizes are $w_{out}=1$ (no sliding window), but their input is kept $w_{in}>1$. For the latter case, the models estimate the most recent time bin at the output.

\subsection{Loss function}

According to Eq.~\ref{eq:Shat}, the estimated ratio mask $\hat{M}(t,f)$ is applied to the noisy speech as $\hat{S}(t,f) = \hat{M}(t,f) \odot Y(t,f)$, the result of which is used for computing the training loss. The loss function considered in training is the compressed mean square error for magnitude \cite{Braun2021Consolidated} $||\hat{S}(t,f)^{0.3} - S(t,f)^{0.3}||^{2}_{2}$, since it has been demonstrated that higher speech quality metrics can be achieved with the use of compression.

For the case where the output has only one time bin, the loss is calculated as the previously mentioned loss function, while for the sliding window case, the loss function is modified to $||\hat{S}(w,t,f)^{0.3} - S(w,t,f)^{0.3}||^{2}_{2}$, in which $\hat{S}(w,t,f)$ is a tensor with the denoised speech STFT for $w$ frames, and $S(w,t,f)$ is the corresponding clean speech target.

\vspace{\spacebeforesecs}
\section{Numerical experiments}

\label{sec:experiments}


\begin{table*}[!t]
    \centering
    \caption{STOI and PESQ averaged over the test set for the noisy signal and denoised signal by the CRN and CDAE model at -5, 0, 10, and 20 dB SNR for different context window sizes. $w_{in} = w_{out} \neq 1$ represent the sliding window scenario. The parameters increase in relation to the $w_{in}=w_{out}=1$ (baseline) for both models case is shown in \%.}
    \begin{tabular}{c c c c c c c c c c c}
        \hline
        
        & & \multicolumn{2}{c}{-5 dB SNR} & \multicolumn{2}{c}{0 dB SNR} & \multicolumn{2}{c}{10 dB SNR} & \multicolumn{2}{c}{20 dB SNR} & Parameters\\
        
        \textbf{Model} & \textbf{Context window} & \textbf{STOI} & \textbf{PESQ} & \textbf{STOI} & \textbf{PESQ} & \textbf{STOI} & \textbf{PESQ} & \textbf{STOI} & \textbf{PESQ} & increase (\%) \\
        
        \hline
        
        & \gr{noisy signal} & \gr{0.679} & \gr{1.360} & \gr{0.767} & \gr{1.505} & \gr{0.907} & \gr{2.053} & \gr{0.974} & \gr{2.914} & \\

        \hline
        
        \multirow{7}{*}{CRN} & $w_{in}=1$, $w_{out}=1$ & 0.781 & 1.674 & 0.860 & 2.000 & 0.949 & 2.878 & 0.983 & 3.689 & baseline \\
        \arrayrulecolor{gray}\cline{2-11}
        & $w_{in}=3$, $w_{out}=1$ & 0.784 & 1.687 & 0.861 & 2.014 & 0.950 & 2.904 & 0.983 & 3.712 & 0.016 \\
        & $w_{in}=3$, $w_{out}=3$ & \red{0.789} & \red{1.698} & \red{0.865} & \red{2.025} & \red{0.951} & \red{2.908} & \red{0.984} & \red{3.713} & 0.032 \\    
        \arrayrulecolor{gray}\cline{2-11}
        & $w_{in}=8$, $w_{out}=1$ & 0.781 & 1.709 & 0.858 & 2.033 & 0.949 & 2.888 & 0.982 & 3.683 & 0.055 \\
        & $w_{in}=8$, $w_{out}=8$ & \red{0.794} & \red{1.730} & \red{0.868} & \red{2.076} & \red{0.952} & \red{2.960} & \red{0.984} & \red{3.763} & 0.112 \\ 
        \arrayrulecolor{gray}\cline{2-11}
        & $w_{in}=13$, $w_{out}=1$ & 0.782 & 1.673 & 0.858 & 2.003 & 0.950 & 2.890 & 0.983 & 3.710 & 0.095 \\
        & $w_{in}=13$, $w_{out}=13$ & \red{0.800} & \red{1.734} & \red{0.868} & \red{2.070} & \red{0.952} & \red{2.950} & \red{0.984} & \red{3.765} & 0.190 \\     

        \hline		
        \hline

        \multirow{3}{*}{CDAE} & $w_{in}=1$, $w_{out}=1$ & 0.696 & 1.390 & 0.786 & 1.593 & 0.899 & 2.248 & 0.946 & 2.992 & baseline \\
        \arrayrulecolor{gray}\cline{2-11}
        & $w_{in}=8$, $w_{out}=1$ & 0.709 & 1.428 & 0.800 & 1.662 & 0.914 & 2.393 & 0.958 & 3.176 & 0.517 \\
        & $w_{in}=8$, $w_{out}=8$ & \red{0.733} & \red{1.500} & \red{0.817} & \red{1.759} & \red{0.920} & \red{2.540} & \red{0.962} & \red{3.392} & 1.038 \\     
        \hline
    \end{tabular}
    \label{tab:results}
\end{table*}

The experiments were carried on with the clean speech files from LibriTTS dataset \cite{Zen2019Libritts}, which is a derivation of the LibriSpeech dataset, and with the noise files from the 5\textsuperscript{th} Deep Noise Suppression Challenge dataset of ICASSP 2023 \cite{Dubey2023icassp}, which is composed of various types of environment noise. The goal is to obtain a speech enhancement NN model generalized over various background noise conditions and signal-to-noise ratios (SNRs). 

\subsection{Data augmentation}
\label{ssec:data}

All audio files are initially resampled from 24~kHz to 16~kHz. Then, for the duration of a randomly drawn noise file (10~s), clean speech segments are randomly drawn and concatenated until the 10~s time is reached, where each clean speech trace receives a gain, randomly chosen, in the range of -3 to 3~dB. Fade in and fade out\footnote{The fade in/out curve is obtained as $0.001e^{6.908 \cdot t}$, which allows for a dynamic range of 60~dB.}, randomly chosen within 0.20 and 0.30~s, is applied to every clean speech trace, as well as for every noise file. Both 10~s noise and clean speech files are normalized by their maximum amplitude, and combined with a uniformly randomly chosen SNR within -5 and 20~dB. The noise, the clean speech, and the noisy speech are then converted to the (STFT) time-frequency domain, with the following parameters: frame length of 128 samples (8~ms), frame hop of 64 samples (4~ms), and a Hanning window function. For training, 100 hours of audio from the LibriTTS 'train-clean-100' subset are considered, while approximately 5.6 hours (1944 files) are used for testing from the 'test-clean' subset. The noise files are randomly (without repetition) drawn from the noise set of the DNS dataset.

\subsection{Results}

For a principled investigation aimed at evaluating the influence of the time-context window size in the STOI and PESQ metrics, we started with the CRN model as a baseline. We trained the model for input windows of size $w_{in}=1,\ 3,\ 8,\ 13$, and two cases for the output: 1) $w_{out} = w_{in}$; and 2) $w_{out}=1$. Notice that the second scenario was considered in previous work \cite{Wang2014OnTraining}, showing a small but consistent increase in speech quality and intelligibility metrics, which we take into account for a more fair comparison against the proposed method. 

The trained models are evaluated in terms of STOI and PESQ with the test dataset as described in Sec.~\ref{ssec:data}. The first part of Table~\ref{tab:results} shows the obtained performance metrics for the noisy signal, as well as the CRN-denoised audio with no context window ($w_{in},w_{out} = 1$), a context window in the input ($w_{in} \neq 1, w_{out}=1$), and context windows in the input and output -- sliding window post-processing case ($w_{out}=w_{in} \neq 1$). Note that the values shown in the table are the average over the entire test dataset. For all cases with the CRN model, the training was executed for 50 epochs and the batch size was set to 32. We used the Adam optimizer with a learning rate of $5 \cdot 10^{-4}$.

Using the insights from our baseline model, we proceed to evaluate the performance of the CDAE architecture. In this case, the considered window sizes for the input were $w_{in}=1,\ 8$ and two cases for the output as considered before: 1) $w_{out} = w_{in}$; and 2) $w_{out}=1$. The training procedure was kept the same as for the baseline model, with the only adjustment being that 40 epochs were used, because of the faster saturation of the loss in training.

We see that, for the CRN model, the use of an explicit time-context window enhanced PESQ and STOI in all cases. Generally, the bigger the size of the context window, the bigger is the increase in the intelligibility and signal quality. Nevertherless, this is expected to saturate for large windows, which however are infeasible due to the introduced processing delay. The use of the sliding window post-processing ($w_{out}=w_{in} \neq 1$) achieved the best results. Because of the convolutional architecture, the number of parameters is only slightly affected (0.06\% increase for $w_{in} = 3$) by the use of the context window, making such an approach a viable post-processing option for such classes of models. It is worth noting that for feedforward and recurrent architectures where the subsequent hidden layers' sizes are increased as the input size increases, the sliding window method will entail higher complexity. Moreover, it is worth mentioning that although the degree of increase in STOI and PESQ metrics is small, a similar increase is obtained when a CRN model is compared to its deep filter version in \cite{schroter2022deepfilternet}.

Our results confirmed that the use of an explicit time-context window is also effective for the CDAE model, providing consistent gains in all examined cases. We observe that the application of the method was more effective in enhancing STOI and PESQ metrics obtained with the CDAE than with the CRN model. This is due to the lack of temporal processing -- which originates from the lack of recurrent or temporal layers in the architecture -- present in the CDAE, which is purely brought by the post-processing. 

It can be envisaged that, due to the explicit inherent temporal structure of architectures such as the temporal convolutional networks (TCN), the sliding window technique could be less effective if applied to such a model class.

\vspace{\spacebeforesecs}
\section{Conclusion}

We investigated the improvement in speech enhancement performance achieved by the use of an explicit time-context window. We examined the application of the method in convolutional recurrent as well as purely convolutional autoencoder architectures. Our results show that the technique can be beneficial at a wide range of SNRs, with highest gains achieved at the most challenging (noisy) cases. For the considered window ranges, a wider window brought more increase in STOI and PESQ metrics. For convolutional-based models, the use of the time-context window only slightly increased the complexity in terms of number of parameters, making it suitable for application in such models. For future work, extension to learnable weighted averaging in the mask estimation as well as extension to complex masks could be considered.

\vfill\pagebreak
\FloatBarrier
\newpage

\vfill\pagebreak
\FloatBarrier

\bibliographystyle{IEEEbib}
\bibliography{refs}

\end{document}